\documentclass[letterpaper, 10 pt, journal, twoside]{IEEEtran}  

\pdfoutput=1 
\usepackage{graphics,import} 
\usepackage{epsfig} 
\usepackage{mathptmx} 
\usepackage{times} 
\usepackage{amsmath} 
\usepackage{amssymb}  
\usepackage{subcaption}
\usepackage{xcolor}
\usepackage{bm}
\usepackage{booktabs}
\usepackage{threeparttable}
\usepackage{tcolorbox}
\usepackage[font={footnotesize}]{caption}
\usepackage{url}

\usepackage[normalem]{ulem}

\title{Dynamic Density Estimation in Heterogeneous Cell Populations}
\author{ \parbox{6in}{\centering Armin K{\"u}per*\thanks{* Corresponding author: armin.kuper@kuleuven.be}, Robert D{\"u}rr, and Steffen Waldherr\\
        Bio- \& Chemical Systems Technology, Reactor Engineering and Safety (CREaS)\\
        Department of Chemical Engineering\\
        KU Leuven\\
        3001 Leuven, Belgium\\
        {\tt\small \{armin.kuper,steffen.waldherr\}@kuleuven.be}, {\tt\small duerr@mpi-magdeburg.mpg.de}}\\
        
}

\newcommand{\MYfooter}{\smash{\scriptsize
\hfil\parbox[t][\height][t]{\textwidth}{\centering
2475-1456 (c) 2018 IEEE. Personal use is permitted, but republication/redistribution requires IEEE permission. See $http://www.ieee.org/publications_standards/publications/rights/index.html$ for more information.}\hfil\hbox{}}}

\newcommand{\MYheader}{\smash{\scriptsize
\hfil\parbox[t][\height][t]{\textwidth}{\centering
This article has been accepted for publication in a future issue of this journal, but has not been fully edited. Content may change prior to final publication. Citation information: DOI 10.1109/LCSYS.2018.2847905, IEEE Control Systems Letters}\hfil\hbox{}}}

\makeatletter

\def\ps@headings{%
\def\@oddhead{\MYheader}
\def\@evenhead{\MYheader}
\def\@oddfoot{\MYfooter}%
\def\@evenfoot{\MYfooter}}

\def\ps@IEEEtitlepagestyle{%
\def\@oddhead{\MYheader}%
\def\@evenhead{\MYheader}%
\def\@oddfoot{\MYfooter}%
\def\@evenfoot{\MYfooter}}

\makeatother

\begin{document}
\maketitle
%
%
\begin{abstract}
%
Multicellular systems play a key role in bioprocess and biomedical engineering.
Cell ensembles encountered in these setups show phenotypic variability like size and biochemical composition.
As this variability may result in undesired effects in bioreactors, close monitoring of the cell population heterogeneity is important for maximum production output, and accurate control.
However, direct measurements are mostly restricted to a few cellular properties.
This motivates the application of model-based online estimation techniques for the reconstruction of non-measurable cellular properties.
Population balance modeling allows for a natural description of cell-to-cell variability.
In this contribution, we present an estimation approach that, in contrast to existing ones, does not rely on a finite-dimensional approximation through grid based discretization of the underlying population balance model.
Instead, our so-called characteristics based density estimator employs sample approximations.
With  two and three-dimensional benchmark examples we demonstrate that our approach is superior to the grid based designs in terms of accuracy and computational demand.
\end{abstract}
%
%
%
\section{INTRODUCTION}
%
\IEEEPARstart{M}{ulticellular} systems are a fundamental part of biomanufacturing processes and typically display phenotypic cell-to-cell variations, for instance in cell size, gene expression levels or concentrations of intracellular signaling molecules.
This heterogeneity affects not only individual cell dynamics but also the interaction of cells with their kind and other species, and is presumed to cause reduced product yields and even process instabilities in bioreactors \cite{Binder2017}.
Close monitoring of the cell population's dynamics is therefore important for efficient biomanufacturing.
\par
Experimental data of heterogeneous cell populations is available through high-throughput single cell measurements (e.g. flow cytometry) in the form of population snapshot data \cite{Hasenauer2011}.
This setup does not provide single cell time series data but information on a representative sample from the population at discrete time points.
The samples can be represented by density distributions with respect to the measured cellular properties. 
Examples of the collection of such data to analyze population heterogeneity are \cite{muller1997dynamics}, where three cellular properties of yeast are measured, or \cite{zuleta2014dynamic}, where an automated flow cytometry setup is presented that can capture population snapshots with high time resolution.
Technical and financial restrictions usually prevent the direct measurement of all relevant intracellular states. To reconstruct the non-measurable quantities, model-based online state estimation methods are required. Here, available data is continuously combined with mathematical model predictions.
\par
Population balance modeling is a framework to account for heterogeneity in large, dynamic populations that is also well established for describing populations of living cells \cite{ramkrishna2014population}.
The resulting population balance equation (PBE) is a multi-dimensional partial differential equation (PDE) that describes the dynamics of the cell distribution through a number density function (NDF) with respect to the distributed cellular properties.
State-of-the-art methods in estimation and control for PBE models typically use grid based approaches and are mostly restricted to one- or two-dimensional density functions, as for example encountered in chemical engineering setups \cite{Mangold2012,porru2017monitoring}.
The computational load increases exponentially with the model dimension for these grid based approaches, rendering them infeasible for high-dimensional PBEs that often arise for multicellular systems.
\par
Motivated by this, we introduce a so-called characteristics based density estimator.  Rather than solving the PBE through a finite-dimensional
approximation, we use a sample approximation by drawing a set of candidate cells from the initial distribution.
Candidate cells are propagated with the single cell model. Their distribution is approximated by a Gaussian mixture density (GMD) representing an estimate for the NDF.
Estimated and measured density function are combined through regularized resampling.
This enables an accurate reconstruction of the unmeasured properties with a significantly reduced numerical effort in comparison to the finite-dimensional approximation approach.
%
\subsubsection*{Structure and contributions} We start with an introduction of population balance modeling (Section \ref{sec:hetcellpop}). Subsequently, we explain the fusion of model predictions with measurements on the population level (Section \ref{sec:modelbasedmeas}).
Using multi-dimensional numerical case studies, we show that our proposed density estimator is more accurate and computationally more efficient than a grid based particle filter (Section \ref{sec:numericalcasestudy}).
The numerical case studies resemble pure growth processes. Phenomena such as cell division and death will be incorporated in later works. 
%
\subsubsection*{Related work} For biotechnological setups previous work has already incorporated flow cytometry into the observer and control design, such as in \cite{milias2011silico}.
The underlying model is however an ODE system with states representing average population values and parameters assumed to be scalar, rather than distributed.
The median population fluorescence is used from flow cytometry measurements. Information about the shape of the measured distributions is not exploited.
\par
Flow cytometry data is typically also used in parameter estimation for stochastic reaction network models \cite{munsky2009listening}.
Even though there is a conceptual relation between such models and the PBE models considered in this paper, a crucial difference is that for the stochastic network models, usually a finite-dimensional vector of scalar parameters is estimated \cite{ZechnerRue2012}, whereas in PBE models an infinite-dimensional density function over the parameter (and state) space needs to be estimated.
%
%
\section{Modeling cell populations}
%
\label{sec:hetcellpop}
In classical approaches, ordinary differential equations (ODEs) are applied to characterize the dynamics of individual cells
\begin{equation}
  \label{equ:scm}
  \boldsymbol{\dot{x}} = \boldsymbol{f}(\boldsymbol{x}),\quad \boldsymbol{y} = \boldsymbol{h}(\boldsymbol{x})\,.
\end{equation}
Here, the state vector $\boldsymbol{x} \in \Omega \subset \mathbb{R}^{d_x}$ may consist of internal variables and static parameters.
Usually, the measurement output $\boldsymbol{y} \in \mathbb{R}^{d_y}$ does not contain all cell states: $d_y < d_x$. Cell-to-cell variability can be described with a time-dependent cell number density function $n: \mathbb{R}^{d_x} \times \Omega \rightarrow \mathbb{R}^{+}: (t,\boldsymbol{x}) \mapsto n(t,\boldsymbol{x})$ over the intracellular states \cite{ramkrishna2014population}.
Its temporal evolution can be derived within the framework of population balance modeling and is given by the population balance equation (PBE)
\begin{equation}
  \label{equ:pbe}
  \dfrac{\partial n}{\partial t}(t, \boldsymbol{x}) + \textnormal{div} (n(t,\boldsymbol{x})\boldsymbol{f}(\boldsymbol{x})) = 0\,.
\end{equation}
Here, we assume that the system is well-mixed and therefore neglect spatial derivatives. The second expression on the left hand side describes the evolution of the NDF due to intracellular dynamics.
In this work, we focus on short time horizons and slow population dynamics, where cell division and cell death are negligible. Taking these phenomena into account would result in a non-zero right hand side.
\par
We will impose a no-influx boundary condition $(n\boldsymbol{f})(t,\partial \Omega)=0$. The initial condition is $n(t_0,\boldsymbol{x})=n_0(\boldsymbol{x})$. Given the assumptions above and the boundary condition, the total number of cells $N$ remains constant.
%
%
\par
PBEs represent multi-dimensional PDEs for which analytical solutions are rarely found. Numerical solution schemes based on a finite-dimensional discretization of the distributed variables, e.g. finite volume method \cite{Ferziger2012}, result in a large system of ODEs.
Here, the computational load increases exponentially with the dimension of the PBE. Alternatively, a solution via the method of characteristics \cite{john1978} yields a system of $d_x+1$ ODEs resulting from the reparameterization of the PBE, which has to be solved for the initial and boundary conditions.
Thus, in case of a no-influx boundary condition, the temporal evolution of the NDF can be approximated through a sample approximation of the initial NDF and propagation of each sample through the characteristic ODE system.
Unlike finite-dimensional approximations, the method of characteristics is free of numerical diffusion.
%
%
\section{Model based dynamic density estimation}
\begin{figure}[tb]
  \centering
  \includegraphics[width=1\columnwidth]{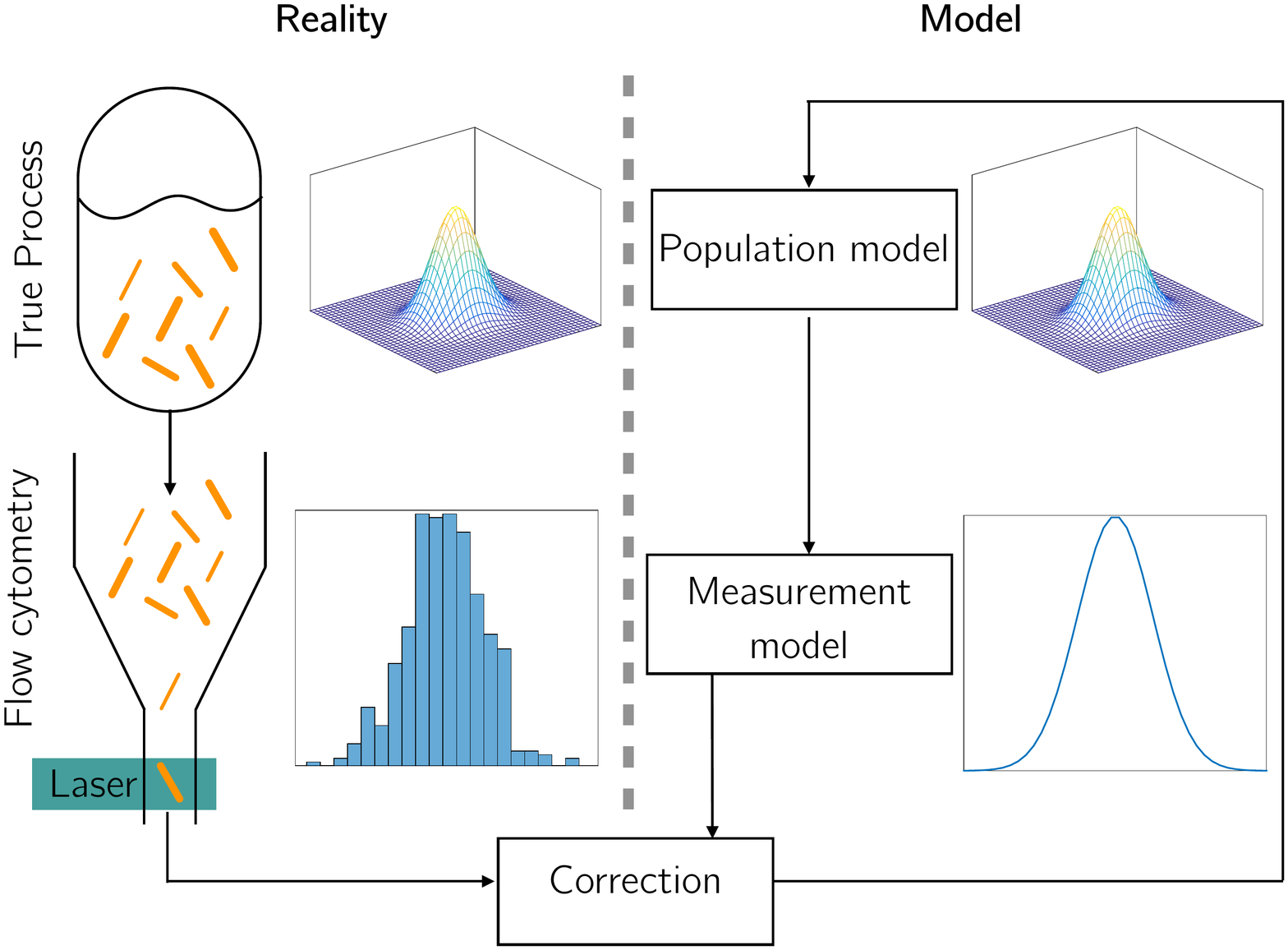}
  \caption{Schematic representation of model based dynamic density estimation in cell population setups. The process model, here a two-dimensional PBE, and a measurement model of flow cytometry are run in parallel to the true process. Prediction and measurement are continuously compared to correct the estimate of the two-dimensional cell heterogeneity.}
  \label{fig:modelbasedmeas}
\end{figure}

%
\label{sec:modelbasedmeas}
Measurements often do not capture the complete cellular state vector $\boldsymbol{x}$. Consider a two-dimensional example where individual cells are heterogeneous with respect to size $z$ and growth rate $g$, thus $\boldsymbol{x}=(z,g)^T$.
Through measurements, only the cell size is available $y=\boldsymbol{h}(\boldsymbol{x}) = z$. Flow cytometry allows the measurement of a vast number of cells which can be represented by a density function $n_y: \mathbb{R}^{d_y} \times \Omega \rightarrow \mathbb{R}^{+}: (t,\boldsymbol{y}) \mapsto n_y(t,\boldsymbol{y})$. This is the NDF marginalized over the non-measurable states and for our two-dimensional example it reads
\begin{equation}
  n_y(t,z)=\int_{0}^{\infty} n(t,z,g)dg.
\end{equation}
With the help of a suitable model to relate measured cell sizes and non-measurable growth rates, the two-dimensional distribution can be reconstructed from the measurements, see Fig~\ref{fig:modelbasedmeas} for a schematic representation of the general idea.
\par
A commonly found paradigm for fusing model predictions with observations is Bayes' theorem \cite{Sarkka2013}. The states and measurements are formulated as probability density functions (PDFs) since they are subject to random perturbations. We will first look at Bayes' rule on the single cell level and later show how it can be applied to the population level. For the single cell model it is
\begin{equation}
  p(\boldsymbol{x}_k | \boldsymbol{y}_{1:k}) = \dfrac{p(\boldsymbol{y}_k | \boldsymbol{x}_k) p(\boldsymbol{x}_k | \boldsymbol{y}_{1:k-1})}{p(\boldsymbol{y}_k | \boldsymbol{y}_{1:k-1})}.
\label{eq:bayes}
\end{equation}
Here, $p(\boldsymbol{x}_k | \boldsymbol{y}_{1:k})$ is the posterior PDF of the state after all measurements until time step $k$ have been factored in. $p(\boldsymbol{y}_k | \boldsymbol{x}_k)$ is the likelihood function and describes how likely the  measurement for the predicted state is. $p(\boldsymbol{x}_k | \boldsymbol{y}_{1:k-1})$ is the prior predicted PDF of the state, which we can obtain from our process model. Finally, $p(\boldsymbol{y}_k | \boldsymbol{y}_{1:k-1})$ is a normalizing constant. There are different methods to solve Bayes' theorem, such as Kalman filtering, or particle filtering. We will briefly recapitulate particle filtering in the next section and then apply it to the PBE.
%
\subsection{Particle filtering}
%
Particle filters solve Bayes' theorem \eqref{eq:bayes} through sequential Monte Carlo sampling. A set of $N_P$ samples (particles) $\boldsymbol{x}_k^{i}$ and weights $w_k^{i}$ with $i=1,...,N_P$ is used to approximate the posterior PDF. Through sequential adjustments of the weights according to
\begin{equation}
\label{eq:weightupdate}
  w_k^{i} \propto \dfrac{ p(\boldsymbol{y}_k|\boldsymbol{x}_k^{i}) p(\boldsymbol{x}_k^{i} | \boldsymbol{x}_{k-1}^{i})}{\pi (\boldsymbol{x}_k^{i} | \boldsymbol{x}^{i}_{0:k-1}, \boldsymbol{y}_{1:k})} w^{i}_{k-1}\,
\end{equation}
the posterior PDF is approximated. Here, a proposal distribution $\pi(\cdot)$ is used instead of the normalizing constant $p(\boldsymbol{y}_k | \boldsymbol{y}_{1:k-1})$, as sampling from the latter is often not possible. Above we assumed that the model has Markov properties. The weights are normalized such that they sum to unity.
We are free in choosing the importance density $\pi (\cdot)$. An obvious choice is to set the importance density as the prior density $p(\boldsymbol{x}_k^{i} | \boldsymbol{x}_{k-1}^{i})$. This is called a bootstrap particle filter. Now, only the likelihood function has to be evaluated to update the weights. 
\par
A common issue with particle filters is that the particle set degenerates, so that only a few particles will have a significant weight \cite{Sarkka2013}. This can be overcome by resampling a new set of particles around those that have a significant weight. Simply copying high weight particles will lead to an impoverishment, i.e. all particles will be identical. Regularized resampling \cite{Sarkka2013} prevents this by forming a continuous distribution of the particle-weight set.
\par
Resampling every time instance is not always needed. We therefore introduce an effective number of particles $N_{eff} \approx 1/(\sum_{i=1}^{N_P} (w_k^{i})^2)$ \cite{Sarkka2013}. 
%
%
\subsection{Particle filter for the discretized PBE}
%
Discretizing the PBE \eqref{equ:pbe} on a grid with $N_{node}$ nodes yields a large system of ODEs which we can formulate as a state space model
\begin{equation}
  \dfrac{d \boldsymbol{n}^{grid}}{dt} = \boldsymbol{A} \boldsymbol{n}^{grid} + \boldsymbol{r}^{grid}\,.
\end{equation}
Here, $\boldsymbol{A} \in \mathbb{R}^{N_{node} \times N_{node}}$ is a sparse matrix resulting from the discretization of the divergence term div$\left(n(t,~\boldsymbol{x})~ \boldsymbol{f}(\boldsymbol{x}) \right)$. The NDF is now represented as a vector $\boldsymbol{n}^{grid} \in \mathbb{R}^{N_{node} \times 1}$. In the particle filter context, this discretized PBE represents the process equation with process noise $\boldsymbol{r}^{grid}$. The measurement equation is given as
\begin{equation}
  \boldsymbol{n}_{y}^{grid} = \boldsymbol{C} \boldsymbol{n}^{grid} + \boldsymbol{v}^{grid}\,.
  \label{eq:discpbemeas}
\end{equation}
Here, $\boldsymbol{C}$ is the measurement matrix resulting from the discrete approximation of the marginal distribution, and $\boldsymbol{v}^{grid}$ is the measurement noise. Note that the actual measurement noise acts on single cell measurements and not directly onto the density function $n_y$. Therefore, \eqref{eq:discpbemeas} does not give an exact representation of the measurement noise. 
\par
A classical particle filter can now be implemented for the obtained discretized system. The mean of the posterior PDF is given as $\hat{\boldsymbol{n}}_{k|k}^{grid} = \sum_{i=1}^{N_P} w^{i}_k \boldsymbol{n}^{grid,i}_{k|k}$, here in time-discrete form for time step $k$.
%
\subsection{Characteristics based density estimator}
\begin{figure}
\centering
\begin{subfigure}{0.5\columnwidth}
  \centering
  \includegraphics[width=1\columnwidth]{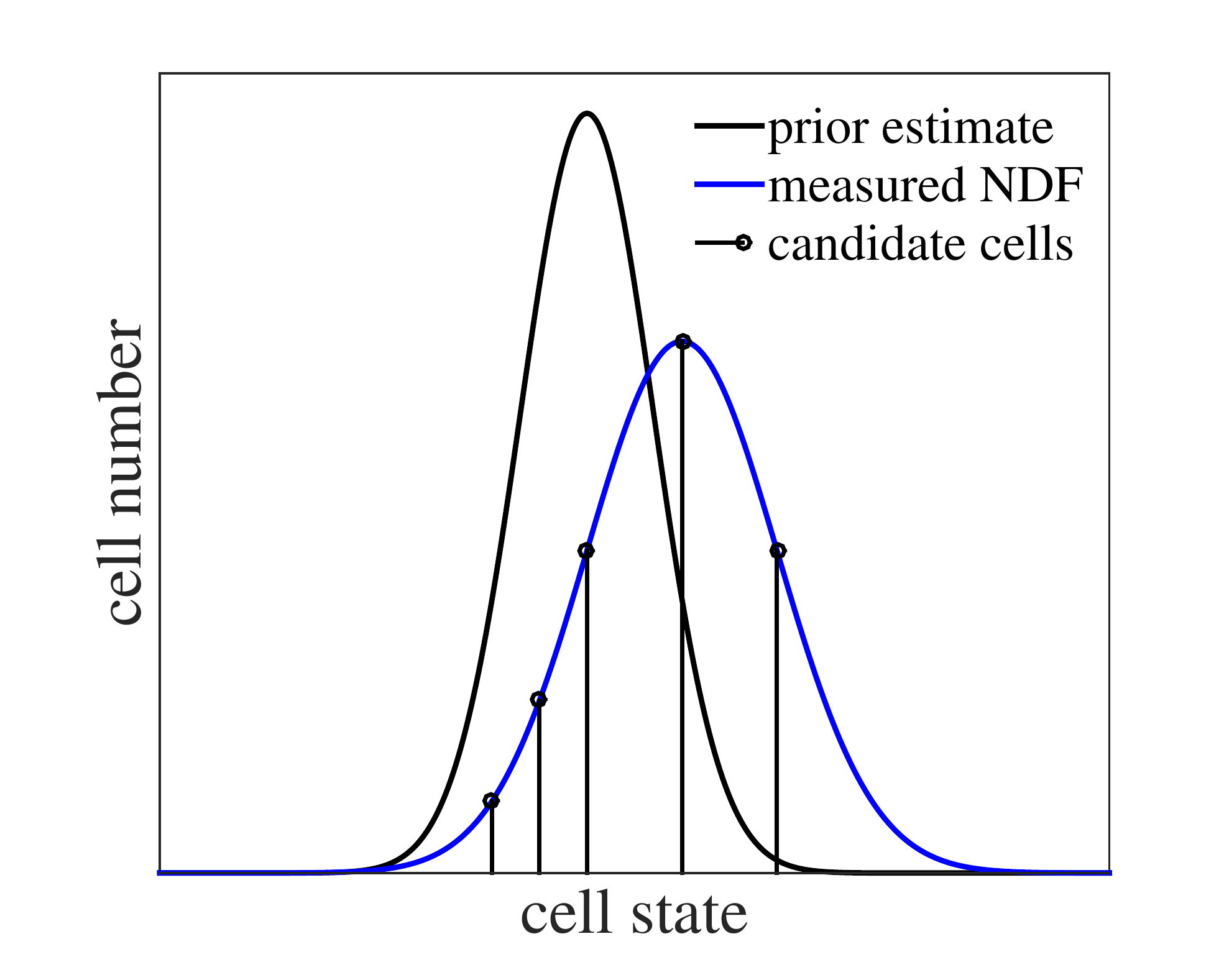}
  \caption{}
  \label{fig:sub1}
\end{subfigure}%
\begin{subfigure}{0.5\columnwidth}
  \centering
  \includegraphics[width=1\columnwidth]{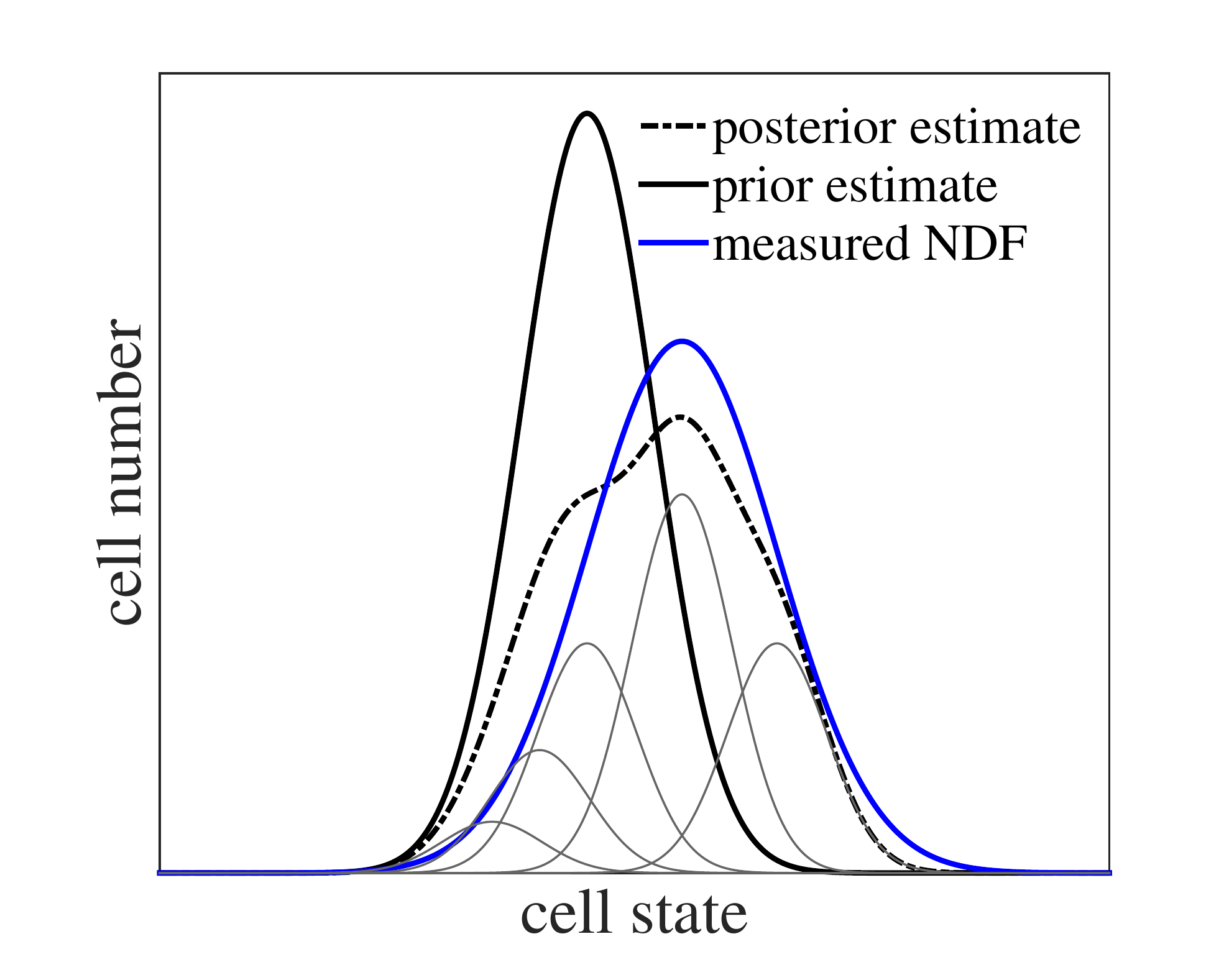}
  \caption{}
  \label{fig:sub2}
\end{subfigure}
\caption{(a) Evaluation of the candidate cells on the measured density function. (b) Combination of estimated and measured density function through the construction of a new GMD.}
\label{fig:regularized_resampling}
\end{figure}

%
The characteristics based density estimator is inspired by the particle filter's sample approximations of the uncertainty PDFs. We apply the sample approximations to the NDF itself. The involved steps are:
\begin{itemize}
  \item[(I)] From the initial NDF $\hat{n}_0 ( \boldsymbol{x} )$ $N_{cand}$ candidate cells $\boldsymbol{x}^{j}$ are sampled. To represent uncertainty of single cells, we attach a covariance $\boldsymbol{W}_0 \in \mathbb{R}^{d_x \times d_x}$ to each candidate cell. We approximate the uncertainty with the unscented transformation \cite{julier2002}, so that for each candidate cell a set of $1 + 2 \cdot d_x$ sigma points is  created. All sigma points of all candidate cells are propagated with the single cell model
  \begin{equation}
    \boldsymbol{\chi}^{j,e}_{k|k-1} = \boldsymbol{f}(\boldsymbol{\chi}^{j,e}_{k-1|k-1}),
  \end{equation}
  with $j=1,...,N_{cand}$ indicating the candidate cell, and $e=0,...,2d_x$ indicating its sigma point.
  \item[(II)] Afterwards, the mean of each sigma point set, its covariance, as well as its corresponding measurement output are calculated
  \begin{align}
    \boldsymbol{x}^{j}_{k|k-1} &= \sum_{e=0}^{2d_x} q^{e}_m \boldsymbol{\chi}^{j,e}_{k|k-1} \\
    \boldsymbol{W}_{k|k-1}^{j} &= \sum_{e=0}^{2d_x} q^{e}_c \left( \boldsymbol\chi_{k|k-1}^{j,e} - \boldsymbol{x}^{j}_{k|k-1} \right) \left( \boldsymbol\chi_{k|k-1}^{j,e} - \boldsymbol{x}^{j}_{k|k-1} \right)^T \\
    \boldsymbol{y}_{k|k-1}^{j} &= \sum_{e=0}^{2d_x} q_m^{e} \boldsymbol{h}\left(  \boldsymbol{\chi}^{j,e}_{k|k-1} \right)
  \end{align}
  Here, $q_c^{e}$ and $q_m^{e}$ are the sigma point weights for the covariance and mean \cite{julier2002}, respectively.
  \item[(III)] The distribution of measured candidate cells is approximated by a GMD
  \begin{equation}
    \boldsymbol{y}_{k|k-1}^{j} \sim \hat{n}^{char}_{y,\,k|k-1} = \sum_{l=1}^{N_{GMD}} \alpha^l_{y} \mathcal{N}(\boldsymbol{y} | \boldsymbol{\mu}_{y}^{l}, \boldsymbol{P}^{l}_{y})\, ,
  \end{equation}
  where $N_{GMD}$ refers to the number of components in the mixture distribution. It should be as low as possible to limit computational cost, while still accurately describing the distribution. Each component $l$ is defined by its mean $\boldsymbol{\mu}_{y}^{l}$, covariance $\boldsymbol{P}^{l}_{y}$, and mixing proportion $\alpha^l_{y}$.
  \item[(IV)] Estimated and measured distribution are now combined by constructing a new GMD. Its mixing proportions are obtained by first evaluating the measured candidate cells on the measurement density function
  \begin{equation}
    \tilde{\alpha}^{j} = n_{y,k}(\boldsymbol{y}_{k|k-1}^{j}),
  \end{equation} 
  and subsequently normalizing $\alpha^{j} = \tilde{\alpha}^{j}/(\sum_j^{N_{cand}} \tilde{\alpha}^{j})$. The posterior GMD is constructed using the candidate cells as the component means
  \begin{equation}
    \boldsymbol{x}^{j}_{k|k} \sim \hat{n}_{k|k}^{char}(\boldsymbol{x}) = \sum_j^{N_{cand}} \alpha^{j} \mathcal{N}(\boldsymbol{x} | \boldsymbol{x}^{j}_{k|k-1}, \boldsymbol{bw}).
  \end{equation}
  This idea is illustrated in Fig \ref{fig:regularized_resampling}. The covariance, or bandwidth $\boldsymbol{bw}$ is calculated via Scott's rule of thumb \cite{Scott2015}, and the same for all candidate cells.
  \item[(V)] Sampling a new set of candidate cells as in (IV) is not always necessary. The Kullback-Leibler divergence \cite{kullback1951information} is used as a measure for the distance between estimated and measured density function
  \begin{equation}
    D_{KL}(\hat{n}_{y,k}^{char} || n_{y,k}).
  \end{equation} 
  We approximate it with a Monte Carlo approximation for which we use the already existing candidate cells as samples. We define a threshold $D_{KL}^{max}$ above which we proceed as in (IV) and reset the covariance $\boldsymbol{W}_0$. Below $D_{KL}^{max}$, we keep the prior candidate cells and the current covariance $\boldsymbol{W}^{j}_{k|k-1}$. 
\end{itemize}
%
\section{Numerical case study}
\label{sec:numericalcasestudy}
\subsubsection*{Two-dimensional benchmark}
The first benchmark process is a two-dimensional PBE that describes a heterogeneous cell population with cells differing in size $z$ and growth rate $g$. Cells first grow with a constant growth rate. After reaching a threshold size $z^{*}=3.5$, their growth rate slowly decreases as they reach a saturation size $z_{max}=6$
\begin{align}
  \label{eq:singlecellgrowth}
  \begin{split}
  \dot{z} =&
  \begin{cases}
  g & \textnormal{if\:} z < z^{*}\\
  g/3.5 \cdot (z_{max} - z) & \textnormal{if\:} z \geq z^{*}
  \end{cases} \\
  \dot{g} =& 0 \qquad y = z.
  \end{split}
\end{align}
Only the cell size is available through measurements. Initial cells are normally distributed
\begin{equation}
  (z_0,\, g_0)^T \sim n_0 = \mathcal{N}(\boldsymbol{\mu}_0, \boldsymbol{P}_0),
\end{equation}
with mean $\boldsymbol{\mu}_0 = (1.5,\, 0.5)^T$ and covariance matrix $\boldsymbol{P}_0= \textnormal{diag}(0.1,\, 0.01)$. The estimators start with an initial estimate of the NDF of $\boldsymbol{\hat{\mu}}_0 = (1.3 \cdot \mu_{1,0},\, 1.3 \cdot \mu_{2,0})^T$ and $\boldsymbol{\hat{P}}_0 = 1.5 \cdot \boldsymbol{P}_0$. The corresponding PBE reads
\begin{align}
  \dfrac{\partial n}{\partial t} (t,z,g)&= - \dfrac{\partial (\dot{z} n)}{\partial z} -   \dfrac{\partial (\dot{g}n)}{\partial g} = -\dfrac{\partial (\dot{z} n)}{\partial z}.
\end{align}
Under the assumption of trivial (no-influx) boundary conditions, we  numerically solve the system with the method of characteristics by solving the initial value problem for 1000 samples drawn from $n_0$.
To mimic experimental snapshot data received through flow cytometry for the cell size, 300 out of the 1000 samples are randomly selected in time steps of $\Delta t = 0.33$ over $t_{sim}=20$.
A bias-free, normally distributed noise with variance $R=0.01$ is added to single cell measurements. Subsequently, a GMD with $N_{GMD}=3$ is fitted to the 300 measured sample cells to construct the measured density function $n_y$.
\subsubsection*{Three-dimensional benchmark}
For the second benchmark process we consider a gene expression model of a heterogeneous cell population. The single cell model reads
\begin{align}
  \dot{z}_1 &= k_1 - z_1\,,\qquad \dot{z}_2 = k_2 z_1 - z_2\,, \qquad
  \dot{k}_1 = 0\,, \nonumber\\
  y &= z_2.
  \label{eq:gene-expression-sc-model}
\end{align}
Cells differ in mRNA concentration $z_1$, protein concentration $z_2$, and transcription rate $k_1$. The translation rate $k_2 = 2$ is assumed to be known and homogeneous throughout the cell population. Only the protein concentration is available through measurements. Initially, cells are distributed according to
\begin{equation}
  (z_{1,0},\, z_{2,0},\, k_{1,0})^T \sim n_0 = \mathcal{N} \left( \boldsymbol{\mu}_0, \boldsymbol{P}_0 \right),
\end{equation}
with mean vector $\boldsymbol{\mu}_0 = (1,\,1,\,2)^T$ and covariance matrix $\boldsymbol{P}_0 =  \boldsymbol{I} \cdot 0.1$, where $\boldsymbol{I} \in \mathbb{R}^{3 \times 3}$. The initial estimate is set to $\hat{\boldsymbol{\mu}}_0 = (1.2 \cdot \mu_{1,0},\, 1.2 \cdot \mu_{2,0},\, 0.8 \cdot \mu_{3,0})^T$ and $\boldsymbol{\hat{P}}_0 = 1.1 \cdot \boldsymbol{P}_0$. Artificial snapshot measurements are generated in the same fashion as for the two-dimensional PBE. In contrast, we now consider noise-free measurements, as well as measurements that are subject to a multiplicative and log-normal distributed noise $\boldsymbol{v}_{\times} \sim \textnormal{exp} \left( \mathcal{N}(\boldsymbol{0}, \boldsymbol{I} \cdot 10^{-2}) \right)$, as this is the main type of disturbance observed in biological experiments \cite{Kreutz2007}.
\subsubsection*{Ensemble observability}
Following the theory developed in \cite{zeng2016ensemble}, ensemble observability denotes the property that the NDF can be reconstructed from the measured marginal distributions.
Applying the conditions discovered in \cite{zeng2016ensemble}, the model~\eqref{eq:gene-expression-sc-model} is ensemble observable.
Because of the non-linearity, no conditions are known to formally check ensemble observability of the first case study model~\eqref{eq:singlecellgrowth}. We conjecture that this is also ensemble observable.
%
%
\subsection{Implementation details}
%
The process equations are implemented in MATLAB in time-continuous form. The ODE solver \textit{ode45} is used for the numerical solution.
The finite volume method is used for the grid based approach. Here, we set the option \textit{NonNegative} to suppress negative values for the solution of the NDF.
Unfortunately, this option increases the computational load drastically. A very fine grid may reduce the magnitude of negative values, but never prevent them completely.
To preserve that sampled particles contain no negative entries for the NDF, we transform the filtering distributions logarithmically, sample the particles, and transform them back again.
\par
For the approximation of sample distributions with a GMD in the characteristics based algorithm, we employed the MATLAB function \textit{gmdistribution.fit} which uses the expectation maximization (EM) algorithm. Tables \ref{tab:pftuning} and \ref{tab:chartuning} list the tuning parameters of a grid based bootstrap regularized resampling particle filter and our characteristics based density estimator for both benchmarks. Ad hoc and process knowledge were used to select appropriate tuning-parameters.
\begin{table}[tb]
\centering
\vspace{8pt}
\caption{\textsc{Tuning-parameters of the grid based particle filter for the two-dimensional case study.}}
\begin{tabular}{@{}r|r@{}}\toprule
 & 2D case study\\ \hline
Particles $N_P$ & 120 \\
Resampling threshold on $N_{eff}$ & $N_P/10$\\
Resampling bw & Scott's rule \\
Nodes in x & 80\\
Nodes in g & 30 \\
\bottomrule
\end{tabular}
\label{tab:pftuning}
\end{table}
\begin{table}[tb]\centering 
\vspace{6pt}
\caption{\textsc{Tuning-parameters of the characteristics based density estimator for the two and three-dimensional case studies.}}
\begin{threeparttable}
\begin{tabular}{@{}r|rr@{}}\toprule
  & \multicolumn{2}{c}{Case study} \\
 & 2D & 3D \\ \hline
 Candidate cells & 300 & 100 \\
 Sampling threshold $D_{KL}^{max} $& 0.08 & 0.05 \\
 Sampling bw & $1/3 \cdot$ Scott's rule  & $3/4 \cdot$ Scott's rule\\
 Cell uncertainty$^{*}$ $\boldsymbol{W}_0$ & $\boldsymbol{I} \cdot 3.86 \cdot 10^{-12}$ & $\boldsymbol{I} \cdot 5.2\cdot 10^{-6}$\\
 GMD components $N_{GMD}$ for $n_y$ & 3 & 3 \\
 EM max. iterations & 500 & 400\\
\bottomrule
\end{tabular}
\begin{tablenotes}\footnotesize
\item[*] The identity matrix $\boldsymbol{I}$ is of dimension $\mathbb{R}^{2 \times 2}$ for the 2D-case and $\mathbb{R}^{3 \times 3}$ for the 3D-case.
\item Tuning-parameters do not differ between the noise-free and noisy three-dimensional case study. 
\end{tablenotes}
\end{threeparttable}
\label{tab:chartuning}
\end{table}
\addtolength{\textheight}{-0.5cm}   

\subsection{Results}
%
\subsubsection*{Two-dimensional benchmark}
\begin{figure}[tb]
  \centering
  \includegraphics[width=1\columnwidth]{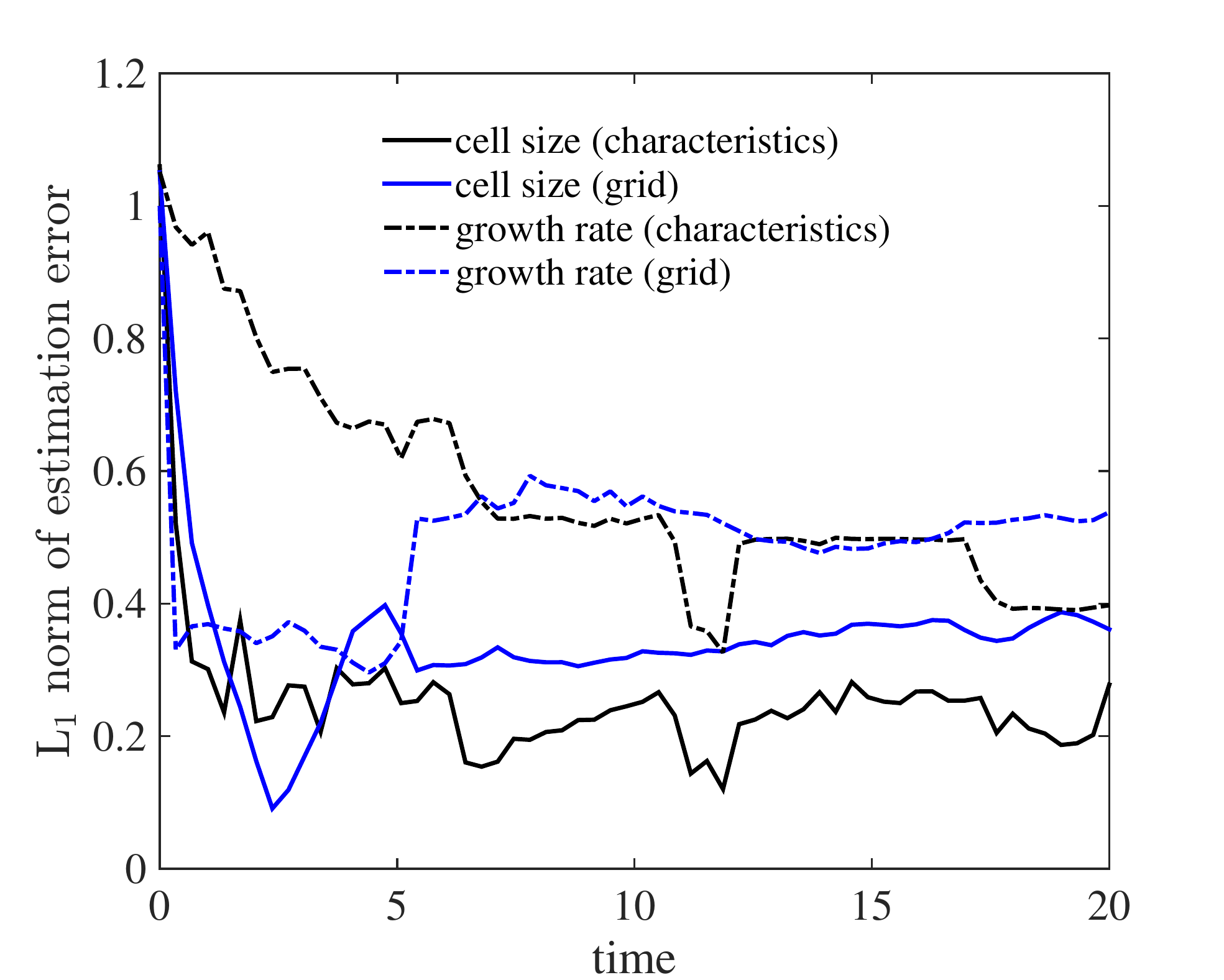}
\caption{Comparison of the L$_1$ norm estimation error for the cell size and growth rate marginal distribution, between the grid based particle filter and characteristics based density estimator.}
  \label{fig:temporal_error_joint_2dpbe}
\end{figure}

We define the L$_1$ norm between estimated and reference marginal NDF as
\begin{equation}
  L_{1,\zeta}(t) = \int_0^{\infty} |n_{\zeta}(t,\zeta) - \hat{n}_{\zeta}(t,\zeta)| d\zeta\,.
\end{equation}
For the cell size and growth rate marginal distribution this is shown in Fig \ref{fig:temporal_error_joint_2dpbe}. For both estimators the estimation error decreases, albeit stronger for the measurable cell size than for the non-measurable growth rate. Note that the initial Gaussian distribution of the cells deforms (not shown) as cells reach the threshold size $z^{*}$ at different time points due to their inherent heterogeneity.
\par
The computational complexity $\mathcal{O}(N_P N_{node}^{d_x})$ increases exponentially with the model dimension $d_x$ for the grid based design. As such, only a few discretization points were employed, see Table \ref{tab:pftuning}. This resulted in numerical diffusion and explains the particle filter's poorer performance. In addition only $120$ particles could be employed.
\par
For the characteristics based design the computational complexity is $\mathcal{O}(d_x N_{cand}(1+2d_x))$ plus the effort of the EM algorithm, which depends on the maximum number of iterations and number of components, see Table \ref{tab:chartuning}. The grid based particle filter algorithm ran for $27.3$ hours on a 6th generation i5 $3.2\, \textnormal{GHz}$ and $8\, \textnormal{GB}$ of RAM, while the characteristics based density estimator was finished in less than $10$ minutes.
\subsubsection*{Three-dimensional benchmark}
For the three-dimensional case study we only implemented the characteristics based approach, as the two-dimensional model was already too computationally demanding for the grid based approach.
\par
\begin{figure*}[tb]
  \centering
  \begin{subfigure}{0.31\linewidth}
    \centering
    \includegraphics[width=1\columnwidth]{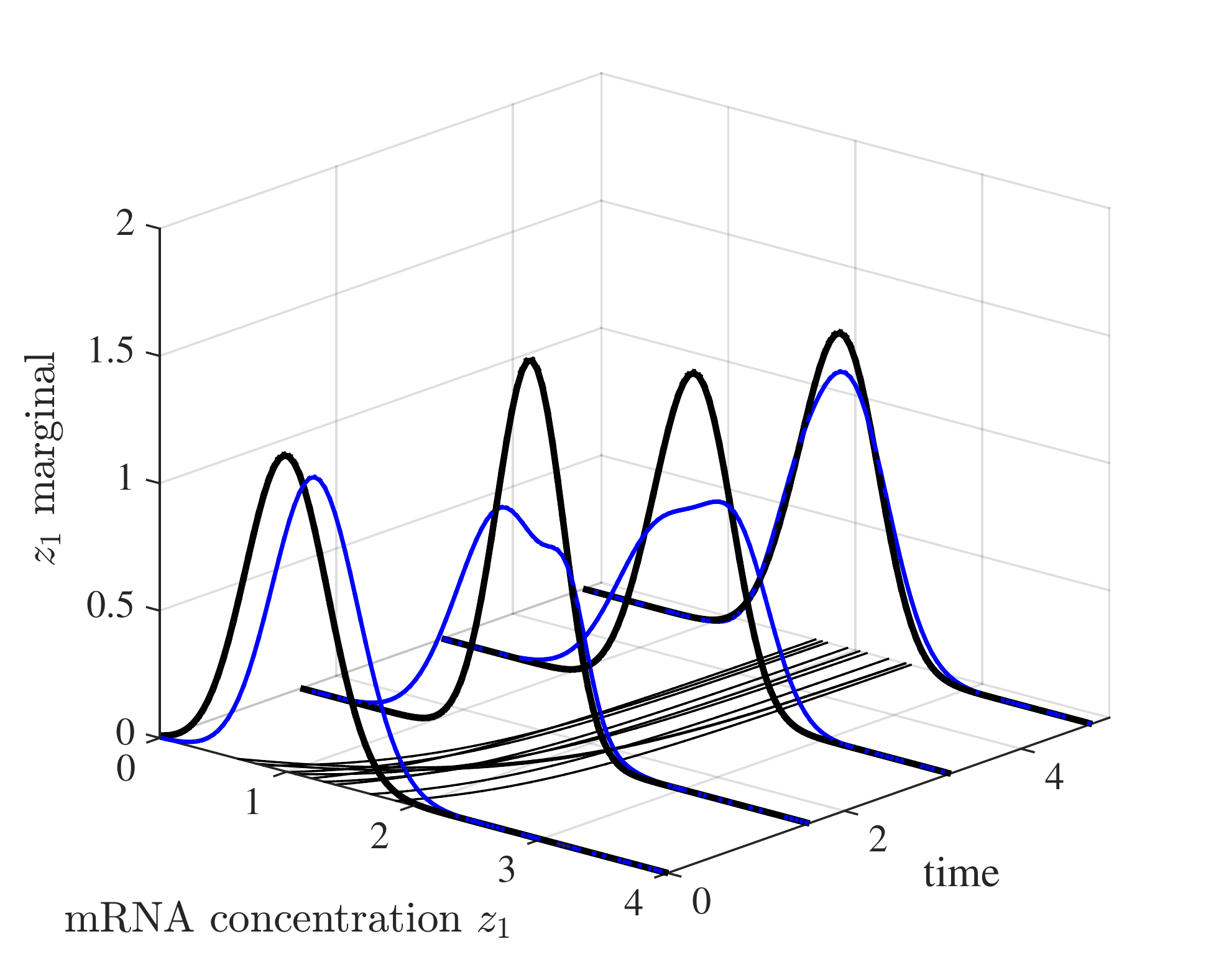} 
    \caption{}
    \label{fig:3dpbe_z1_nonoise}
  \end{subfigure}
  \begin{subfigure}{0.31\linewidth}
    \centering
     \includegraphics[width=1\columnwidth]{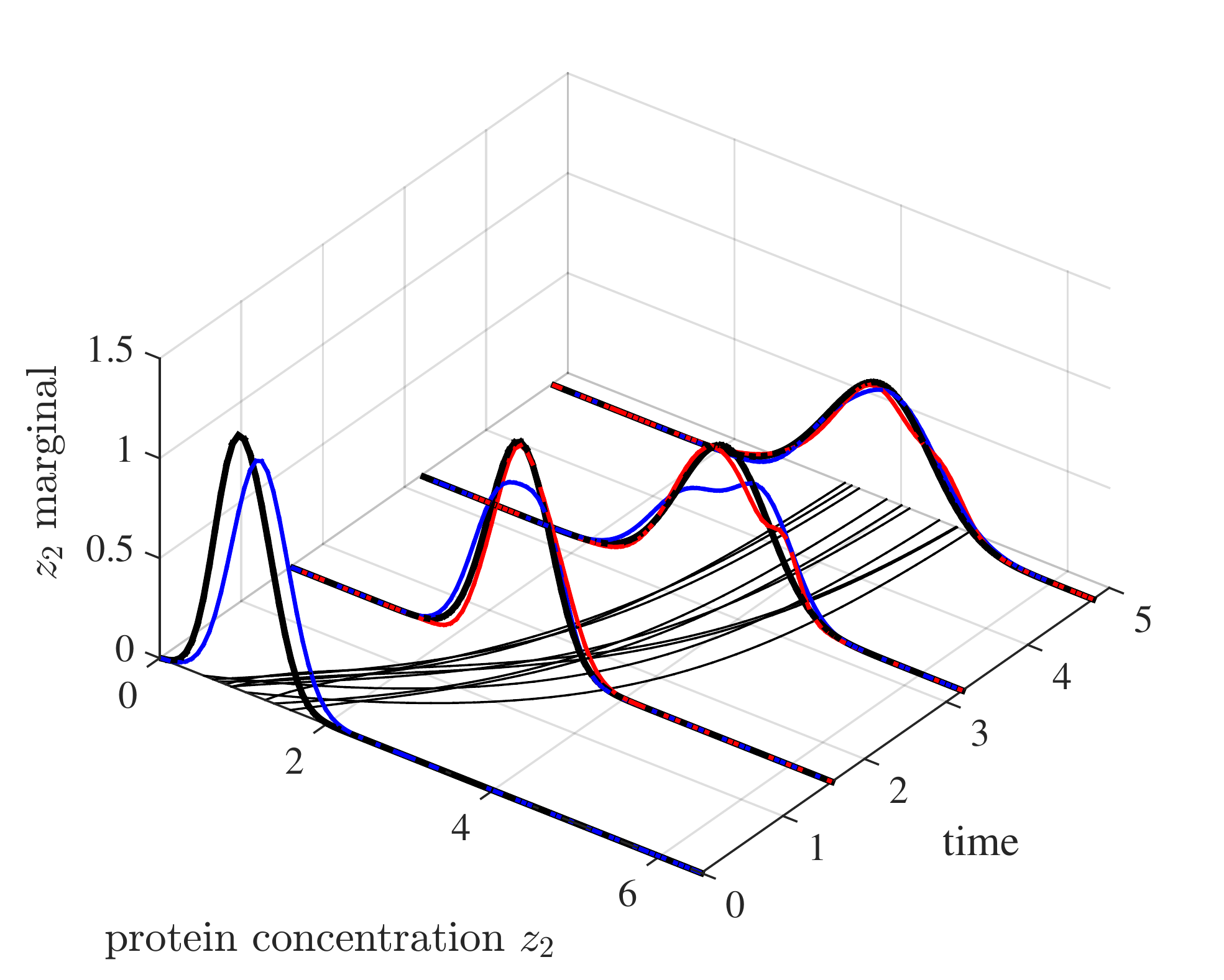} 
    \caption{}
    \label{fig:3dpbe_z2_nonoise}
  \end{subfigure}
  \begin{subfigure}{0.31\linewidth}
    \centering
    \includegraphics[width=1\columnwidth]{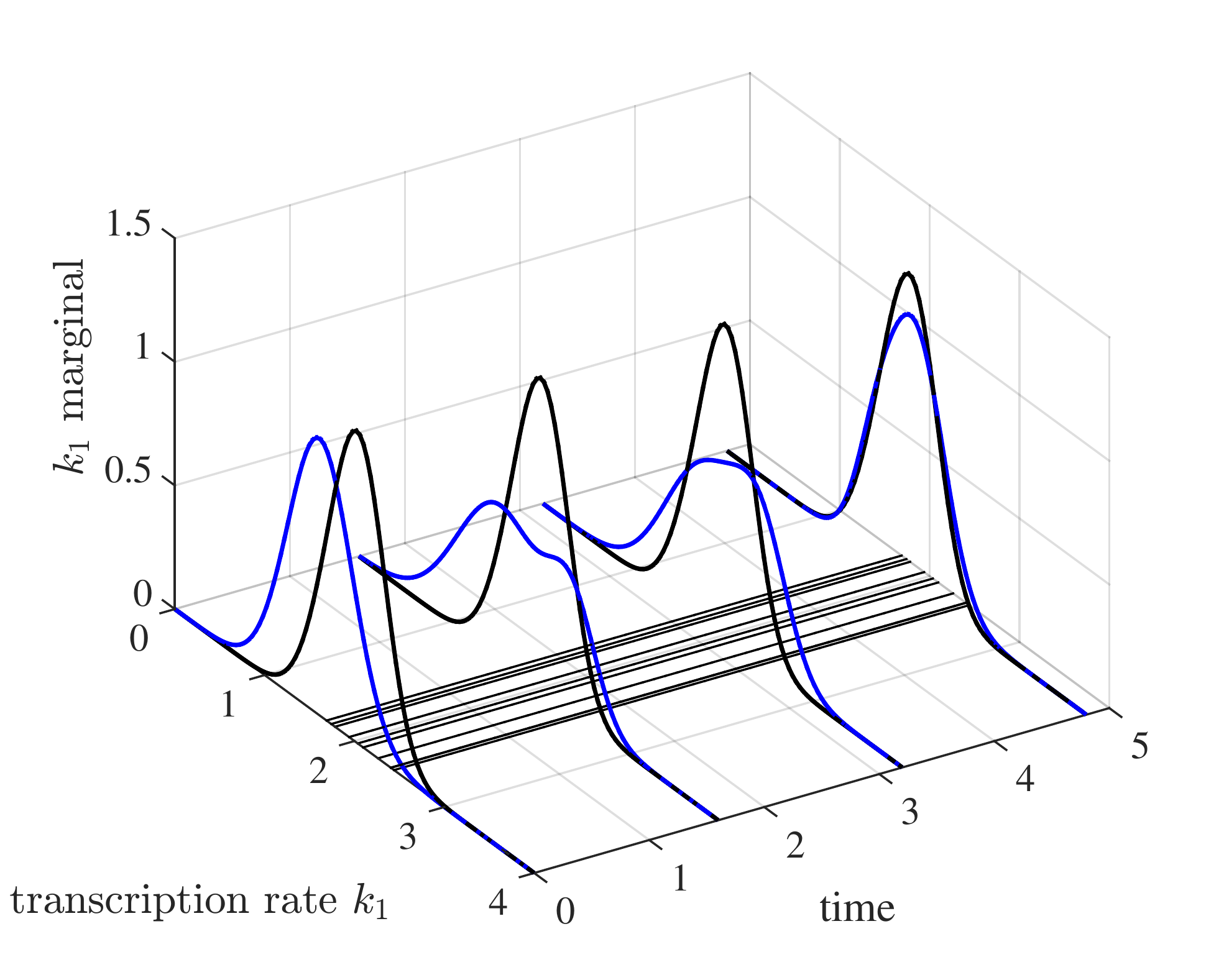} 
    \caption{}
    \label{fig:3dpbe_k1_nonoise}
  \end{subfigure}
  \caption{Development of the NDF at fixed time points. The black line shows the reference NDF, the blue line shows the estimation, and the red line in (a) and (b) shows the measured distribution. The flat, black lines represent the trajectories of a selection of reference cells. All NDFs are normalized by the number of total cells.}
\end{figure*}

The artificial measurements and the estimated marginal distributions of the mRNA concentration $z_1$, the protein concentration $z_2$, as well as the transcription rate $k_1$ are shown in Fig \ref{fig:3dpbe_z1_nonoise}, \ref{fig:3dpbe_z2_nonoise}, and \ref{fig:3dpbe_k1_nonoise}, respectively. All distributions are normalized by the total number of cells. Our characteristics based approach is able to track the measured, as well as the unmeasured marginal distribution with sufficient accuracy. Remarkably, the artificial measurement distribution differs from the reference distribution, although measurements on the single cell level are considered to be noise-free. The nature of the measurement setup reveals why: it only allows to obtain a sample of the cell population, which might not accurately represent the whole population.
\par
\begin{figure*}[tb] 
\centering
\begin{subfigure}{0.31\linewidth}
  \centering
   \includegraphics[width=1\columnwidth]{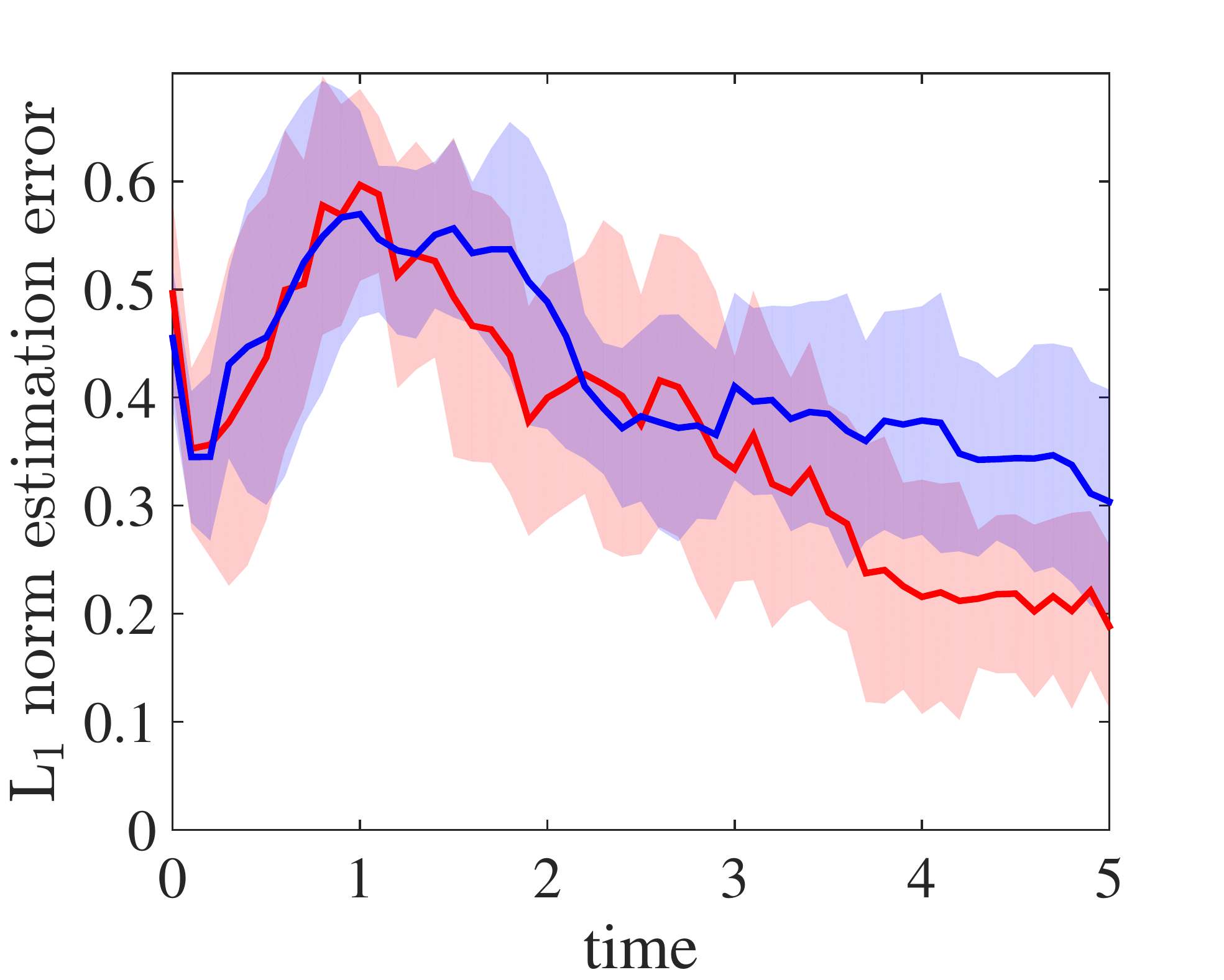} 
  \caption{mRNA concentration $z_1$}
  \label{fig:error_3dpbe_z1}
\end{subfigure}
\begin{subfigure}{0.31\linewidth}
  \centering  
  \includegraphics[width=1\columnwidth]{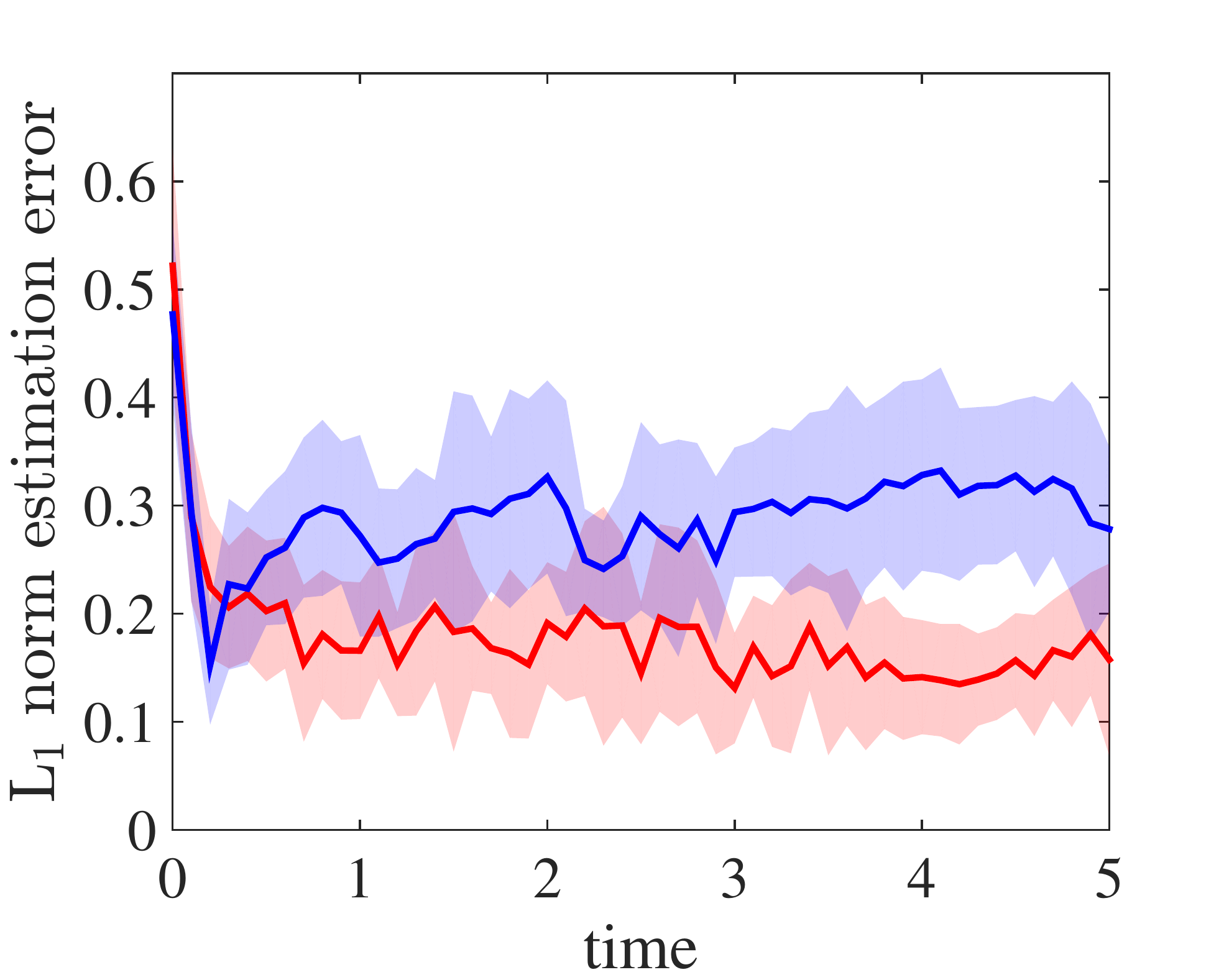} 
  \caption{Protein concentration $z_2$}
  \label{fig:error_3dpbe_z2}
\end{subfigure}
\begin{subfigure}{0.31\linewidth}
  \centering
  \includegraphics[width=1\columnwidth]{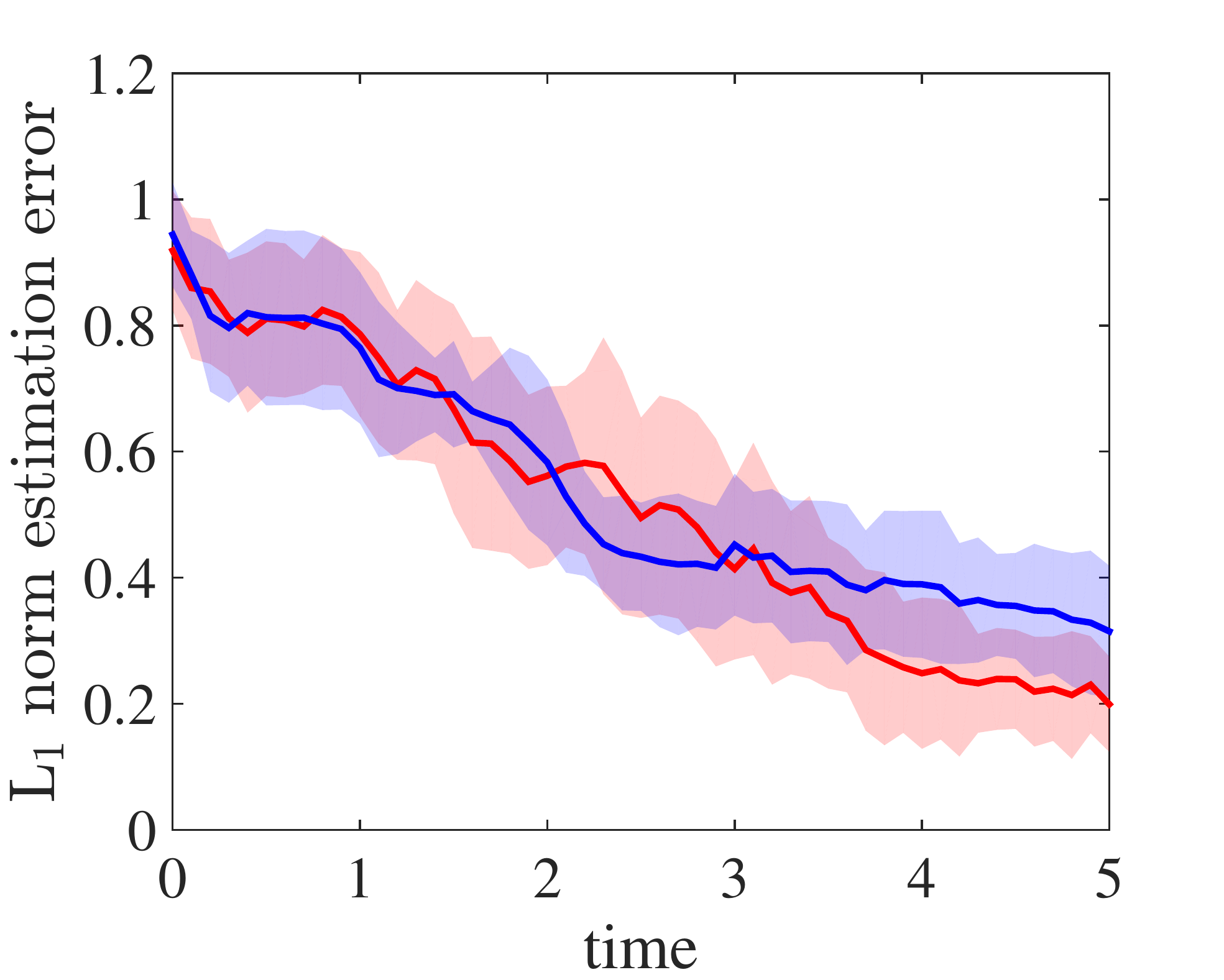}  
    \caption{Transcription rate $k_1$}
    \label{fig:error_3dpbe_k1}
\end{subfigure}
\caption{Estimation error plots for all three states, averaged over ten simulation runs. Red lines show the averaged L$_1$ norm of the estimation error for the noise-free measurements, while blue lines represent the noisy case. Shaded areas indicate one standard deviation from the averaged value.}
\label{fig:error_3dpbe}
\end{figure*}

In addition to the noise-free measurements, we also conducted simulations with measurement noise to examine its influence on estimation performance. Each scenario was simulated 10 times, over which the L$_1$ estimation error was averaged. This is shown in Fig \ref{fig:error_3dpbe} for each marginal distribution. Unsurprisingly, measurement noise negatively affected the estimation performance for all three marginal distributions, though not severely. 
%
%

\section{Summary}
%
We proposed a novel online state estimation approach for the dynamic reconstruction of cell density distributions from cell population snapshot data. In contrast to grid based approaches our characteristics based density estimator uses sample approximations of the NDF. The computational complexity is reduced from scaling exponentially with the model dimension to an approximately quadratic increase with the model dimension. This is a significant improvement, as high-dimensional models often arise from complex cellular dynamics.
We showed for two and three-dimensional benchmarks that our approach is not only less computationally demanding, but also more accurate than the discretization based designs. In future works we plan to extend the proposed technique to problems with cell division and cell death, multimodal distributions, and stochastic single cell dynamics.
%
%
%

\bibliographystyle{IEEEtran}
\bibliography{IEEEabrv,cdc18_lib}

\begin{thebibliography}{10}
\providecommand{\url}[1]{#1}
\csname url@samestyle\endcsname
\providecommand{\newblock}{\relax}
\providecommand{\bibinfo}[2]{#2}
\providecommand{\BIBentrySTDinterwordspacing}{\spaceskip=0pt\relax}
\providecommand{\BIBentryALTinterwordstretchfactor}{4}
\providecommand{\BIBentryALTinterwordspacing}{\spaceskip=\fontdimen2\font plus
\BIBentryALTinterwordstretchfactor\fontdimen3\font minus
  \fontdimen4\font\relax}
\providecommand{\BIBforeignlanguage}[2]{{%
\expandafter\ifx\csname l@#1\endcsname\relax
\typeout{** WARNING: IEEEtran.bst: No hyphenation pattern has been}%
\typeout{** loaded for the language `#1'. Using the pattern for}%
\typeout{** the default language instead.}%
\else
\language=\csname l@#1\endcsname
\fi
#2}}
\providecommand{\BIBdecl}{\relax}
\BIBdecl

\bibitem{Binder2017}
D.~Binder, T.~Drepper, K.-E. Jaeger, F.~Delvigne, W.~Wiechert, D.~Kohlheyer,
  and A.~Gr{\"{u}}nberger, ``{Homogenizing bacterial cell factories: Analysis
  and engineering of phenotypic heterogeneity},'' \emph{Metabolic Engineering},
  vol.~42, pp. 145--156, 2017.

\bibitem{Hasenauer2011}
J.~Hasenauer, S.~Waldherr, M.~Doszczak, N.~Radde, P.~Scheurich, and
  F.~Allg{\"{o}}wer, ``{Identification of models of heterogeneous cell
  populations from population snapshot data.}'' \emph{BMC Bioinformatics},
  vol.~12, no.~1, p. 125, 2011.

\bibitem{muller1997dynamics}
S.~M{\"u}ller, K.~Hutter, T.~Bley, L.~Petzold, and W.~Babel, ``Dynamics of
  yeast cell states during proliferation and non proliferation periods in a
  brewing reactor monitored by multidimensional flow cytometry,''
  \emph{Bioprocess Engineering}, vol.~17, no.~5, pp. 287--293, 1997.

\bibitem{zuleta2014dynamic}
I.~A. Zuleta, A.~Aranda-D{\'\i}az, H.~Li, and H.~El-Samad, ``Dynamic
  characterization of growth and gene expression using high-throughput
  automated flow cytometry,'' \emph{Nature Methods}, vol.~11, no.~4, p. 443,
  2014.

\bibitem{ramkrishna2014population}
D.~Ramkrishna and M.~R. Singh, ``Population balance modeling: current status
  and future prospects,'' \emph{Annual Review of Chemical and Biomolecular
  Engineering}, vol.~5, pp. 123--146, 2014.

\bibitem{Mangold2012}
M.~Mangold, ``{Use of a Kalman filter to reconstruct particle size
  distributions from FBRM measurements},'' \emph{{Chemical Engineering
  Science}}, vol.~70, pp. 99--108, 2012.

\bibitem{porru2017monitoring}
M.~Porru and L.~{\"O}zkan, ``Monitoring of batch industrial crystallization
  with growth, nucleation, and agglomeration. part 1: Modeling with method of
  characteristics,'' \emph{Industrial \& Engineering Chemistry Research},
  vol.~56, no.~20, pp. 5980--5992, 2017.

\bibitem{milias2011silico}
A.~Milias-Argeitis, S.~Summers, J.~Stewart-Ornstein, I.~Zuleta, D.~Pincus,
  H.~El-Samad, M.~Khammash, and J.~Lygeros, ``In silico feedback for in vivo
  regulation of a gene expression circuit,'' \emph{Nature biotechnology},
  vol.~29, no.~12, p. 1114, 2011.

\bibitem{munsky2009listening}
B.~Munsky, B.~Trinh, and M.~Khammash, ``Listening to the noise: random
  fluctuations reveal gene network parameters,'' \emph{Molecular Systems
  Biology}, vol.~5, no.~1, p. 318, 2009.

\bibitem{ZechnerRue2012}
C.~Zechner, J.~Ruess, P.~Krenn, S.~Pelet, M.~Peter, J.~Lygeros, and H.~Koeppl,
  ``Moment-based inference predicts bimodality in transient gene expression,''
  \emph{Proceedings of the National Academy of Sciences}, vol. 109, no.~21, pp.
  8340--8345, 2012.

\bibitem{Ferziger2012}
J.~H. Ferziger and M.~Peric, \emph{Computational methods for fluid
  dynamics}.\hskip 1em plus 0.5em minus 0.4em\relax Springer Science \&
  Business Media, 2012.

\bibitem{john1978}
F.~John, \emph{{Partial Differential Equations}}, 3rd~ed.\hskip 1em plus 0.5em
  minus 0.4em\relax Springer-Verlag, New York, 1978.

\bibitem{Sarkka2013}
S.~S{\"a}rkk{\"a}, \emph{Bayesian filtering and smoothing}.\hskip 1em plus
  0.5em minus 0.4em\relax Cambridge University Press, 2013, vol.~3.

\bibitem{julier2002}
S.~J. Julier, ``The scaled unscented transformation,'' in \emph{Proceedings of
  the 2002 American Control Conference (IEEE Cat. No.CH37301)}, vol.~6, May
  2002, pp. 4555--4559.

\bibitem{Scott2015}
D.~W. Scott, \emph{Multivariate density estimation: theory, practice, and
  visualization}.\hskip 1em plus 0.5em minus 0.4em\relax John Wiley \& Sons,
  2015.

\bibitem{kullback1951information}
S.~Kullback and R.~A. Leibler, ``On information and sufficiency,'' \emph{The
  Annals of Mathematical Statistics}, vol.~22, no.~1, pp. 79--86, 1951.

\bibitem{Kreutz2007}
C.~Kreutz, M.~M.~B. Rodriguez, T.~Maiwald, M.~Seidl, H.~E. Blum, L.~Mohr, and
  J.~Timmer, ``{An error model for protein quantification},''
  \emph{Bioinformatics}, vol.~23, no.~20, pp. 2747--2753, 2007.

\bibitem{zeng2016ensemble}
S.~Zeng, S.~Waldherr, C.~Ebenbauer, and F.~Allg{\"o}wer, ``Ensemble
  observability of linear systems,'' \emph{IEEE Transactions on Automatic
  Control}, vol.~61, no.~6, pp. 1452--1465, 2016.

\end{thebibliography}

\end{document}